\documentclass[twocolumn]{aastex631}

\def\simgr{\,\hbox{\hbox{$ > $}\kern -0.8em \lower 1.0ex\hbox{$\sim$}}\,}
\def\simle{\,\hbox{\hbox{$ < $}\kern -0.8em \lower 1.0ex\hbox{$\sim$}}\,}

\newcommand{\xmm}{XMM-Newton}

\newcommand{\msptwo}{PSR~J1048+2339}
\newcommand{\fgl}{3FGL~J0838.8$-$2829}
\newcommand{\obj}{1FGL~J0523.5$-$2529}
\newcommand{\psr}{4FGL~J0935.3+0901}

\shortauthors{Halpern}
\shorttitle{\psr: A Flaring Black Widow Candidate}

\begin{document}
\title{Optical Light Curve of \psr: A Flaring Black Widow Candidate}

\author[0000-0003-4814-2377]{Jules P. Halpern}
\affiliation{Department of Astronomy and Columbia Astrophysics Laboratory, Columbia University, 550 West 120th Street, New York, NY 10027-6601, USA; jph1@columbia.edu}

\begin{abstract}
  I obtained time-series photometry of the compact binary candidate for the Fermi source \psr.  Superposed on the 2.44~hr orbital modulation are day-to-day variations and frequent flaring as seen in several redback and black widow millisecond pulsars (MSPs).  The short orbital period favors a black widow.   While the modulation of $\leq 1$~mag is smaller than that of most black widows, it could indicate a low orbital inclination.  Although a published optical spectrum shows strong emission lines, the light curve evinces pulsar heating of the companion star rather than accretion-disk emission of a transitional MSP.  Emission lines and flaring occur in the same objects, probably powered by shocks between the relativistic pulsar wind and a wind driven off the companion star.  I also recovered the period in photometry from the Zwicky Transient Facility (ZTF).  A phase-connected ephemeris derived from MDM Observatory and ZTF data spanning 4~yr yields a period of 0.10153276(36)~days and an epoch for the ascending node of the putative pulsar.
\end{abstract}

\section{Introduction\label{sec:intro}}

The standard model for the formation of millisecond pulsars MSPs is spin-up (recycling) of the neutron star by accretion in a low-mass X-ray binary (LMXB) system \citep{alp82,rad82}, followed by turn-on of the radio pulsar mechanism after accretion stops.   Of the 3319 pulsars in version 1.67 of the ATNF catalog\footnote{https://www.atnf.csiro.au/research/pulsar/psrcat/} \citep{man05}, 501 have spin periods $<20$~ms and magnetic dipole field strengths $B_s<10^{10}$~G, a combination that is a reliable indicator of a recycled origin.

The Large Area Telescope on the Fermi Gamma-ray Space Telescope has detected a significant fraction of known and new millisecond pulsars\footnote{https://confluence.slac.stanford.edu/display/GLAMCOG/\\Public+List+of+LAT-Detected+Gamma-Ray+Pulsars}.  A major result is the profusion of new black widow and redbacks MSPs (collectively ``spiders''), previously rare compact binaries having orbital periods $\simle 1$~day, in which the relativistic pulsar wind interacts with a low-mass or substellar companion to heat its photosphere and help drive a wind. Black widows have degenerate companions of $<0.05\,M_{\odot}$, while redbacks have nondegenerate $0.1 - 1\,M_{\odot}$ companions that are hotter and of lower density than main-sequence stars of the same mass.  These were recognized as distinct subclasses when a bimodal mass distribution became apparent among the Fermi MSP identifications \citep{rob11,rob13}.  \citet{hui19} catalogud 44 black widow pulsars (27 in the Galactic field and 17 in globular clusters), and 26 redbacks (14 in the Galactic field and 12 in globular clusters).

Spider companions are close to filling their Roche lobes \citep{dra19,str19}, and their winds often obscure the radio pulsar signal for part of the orbit (e.g., \citealt{cam16,den16,cla21,cor21}), making their timing difficult.  The winds are sometimes seen in optical emission lines, principally H$\alpha$ \citep{hal17b,str19}, but also \ion{He}{1} \citep{rom15}. Nonthermal shock emission from the collision of the pulsar wind and the stellar wind is evident in X-rays \citep{bog05,bog11,bog14,aln18}, and modulated in orbital phase.  Photospheric heating is often strong, attributed to either high-energy photons from the shock or direct bombardment by pulsar wind particles \citep{rom16,san17}.  The optical modulation can be dominated by the contrast between the heated and ``dark'' sides of the companion \citep{bre13,sch14}, or, if the heating effect is small, by ellipsoidal modulation of the light from the tidally distorted star, and starspots \citep{li14,bel16,van16}. All of these effects can change on timescales of minutes to months. Variable heating is seen in the optical light curves over weeks and months, while rapid flares with time scales of minutes increase the luminosity by a factor of 10 or more in X-rays, and up to $\sim1$~mag in optical \citep{hal17a,cho18,an17}.

Some redbacks have an accretion disk producing double-peaked optical emission lines, but one that is truncated by the rotating pulsar magnetosphere or pulsar wind. They also display a unique pattern of ``moding'' in the X-ray and optical \citep{bog15a,bog15b,dem15}.  Known as transitional millisecond pulsars (tMSPs) in the subluminous disk state, the majority were identified from Fermi sources.  tMSPs may be the evolutionary link between the redback MSPs and their LMXB progenitors.  There are additional tMSP candidates whose spin periods are not known because they have not yet been seen in a radio pulsar state.  See \citet{pap22} for a review of tMSPs, their accretion physics, and their relation to redbacks.

The subject of this Letter is the probable counterpart of \psr, which \citet{wan20} identified as a Swift X-ray source and a $2.47\pm0.04$~hr optical photometric binary with a modulation of semiamplitude 0.30~mag.  Strong, double-peaked optical emission lines were also detected.  \citet{wan20} interpreted these properties as consistent with either a redback MSP or a tMSP.

The position of this star is (J2000) R.A.=$09^{\rm h}35^{\rm m}20^{\rm s}\!.719$, decl.=$+09^{\circ}00^{\prime}35^{\prime\prime}\!.90$ in the Gaia extended third data release (EDR3; \citealt{bro21}).  Significant parallax is not detected, although proper motion components are measured: ($\mu_{\alpha}{\rm cos}\,\delta, \mu_{\delta})=(-7.6\pm1.8, -2.9\pm1.1)$~mas~yr$^{-1}$.  Magnitudes in Pan-STARRS $g,r$,and $i$  range from 20--21.5 \citep{wan20}.

\citet{zhe22} followed up with an X-ray light curve and spectrum of \psr\ from \xmm, which they conclude favors a redback in the radio pulsar state rather than a tMSP.  They also conducted a 20 minute pulsar search with the Five-hundred-meter Aperture Spherical radio Telescope (FAST), which did not result in a detection.  However, this observation covered only a small fraction of the orbit at an as yet unknown phase.

Here I present more extensive optical time-series photometry that reveals frequent variability and fast flaring behavior, providing additional evidence of interacting of a pulsar with the companion star.  Section~2 describes the data and derives an ephemeris from time-series photometry at MDM Observatory, supplemented with data from the Zwicky Transient Facility (ZTF; \citealt{bel19}).  Section~3 discusses further implications of these results for the nature of the companion star and its interaction with the putative pulsar.  Conclusions and suggestions for further work are presented in Section~4.

\section{Observations and Analysis}
\subsection{MDM 1.3m\label{sec:obs}}
\begin{deluxetable}{lccrc}
\label{tab:optlog}
\tablecolumns{5} 
\tablewidth{0pt} 
\tablecaption{Log of MDM 1.3~m Time-series Photometry}
\tablehead{
\colhead{Date} & \colhead{Exposure\tablenotemark{a}} &
\colhead{Time} & \colhead{N Exp} & \colhead{Airmass} \\
\colhead{(UT)} & \colhead{(s)} & \colhead{(UTC)} & &
\colhead{(sec $z$)}
}
\startdata
2022 Jan 4  & 200 & 05:46--13:24 & 132 & 1.09--2.30 \\
2022 Jan 5  & 200 & 05:40--13:22 & 132 & 1.09--2.33 \\
2022 Feb 1  & 200 & 06:02--12:02 & 104 & 1.09--1.80 \\
2022 Feb 3  & 200 & 05:55--12:28 & 116 & 1.09--2.23 \\
2022 Feb 9  & 200 & 07:09--10:22 &  57 & 1.09--1.36 \\
2022 Apr 5  & 200 & 02:47--07:55 &  91 & 1.09--1.81 \\
2022 Apr 6  & 200 & 02:42--08:14 &  98 & 1.09--2.07 \\
2022 Apr 20 & 200 & 02:51--05:57 &  55 & 1.09--1.40 \\
\enddata
\tablenotetext{a}{All exposures used a BG38 filter and had 3~s cycle time.}
\end{deluxetable}
I used the MDM Observatory 1.3~m McGraw-Hill telescope for time-series photometry.  The thinned, backside-illuminated CCD ``templeton'' was windowed to achieve an efficient read/prep cycle time of 3~s compared with the 200~s exposure time.  Given the small telescope aperture and faint magnitude of the target, typically in the range 20--21, I used only a single broadband (3200--6500 \AA) BG38 filter to optimize signal to noise, sacrificing absolute calibration and color information.  To create the light curves, differential photometry was performed with respect to a nearby comparison star.  An approximate magnitude was assigned to the comparison star from the average of its $B$ and $R$ in the second Palomar Sky Survey.

\begin{figure*}
\vspace{-2.8in}
\centerline{
\hspace{-0.27in}
\includegraphics[angle=0.,width=1.26\linewidth]{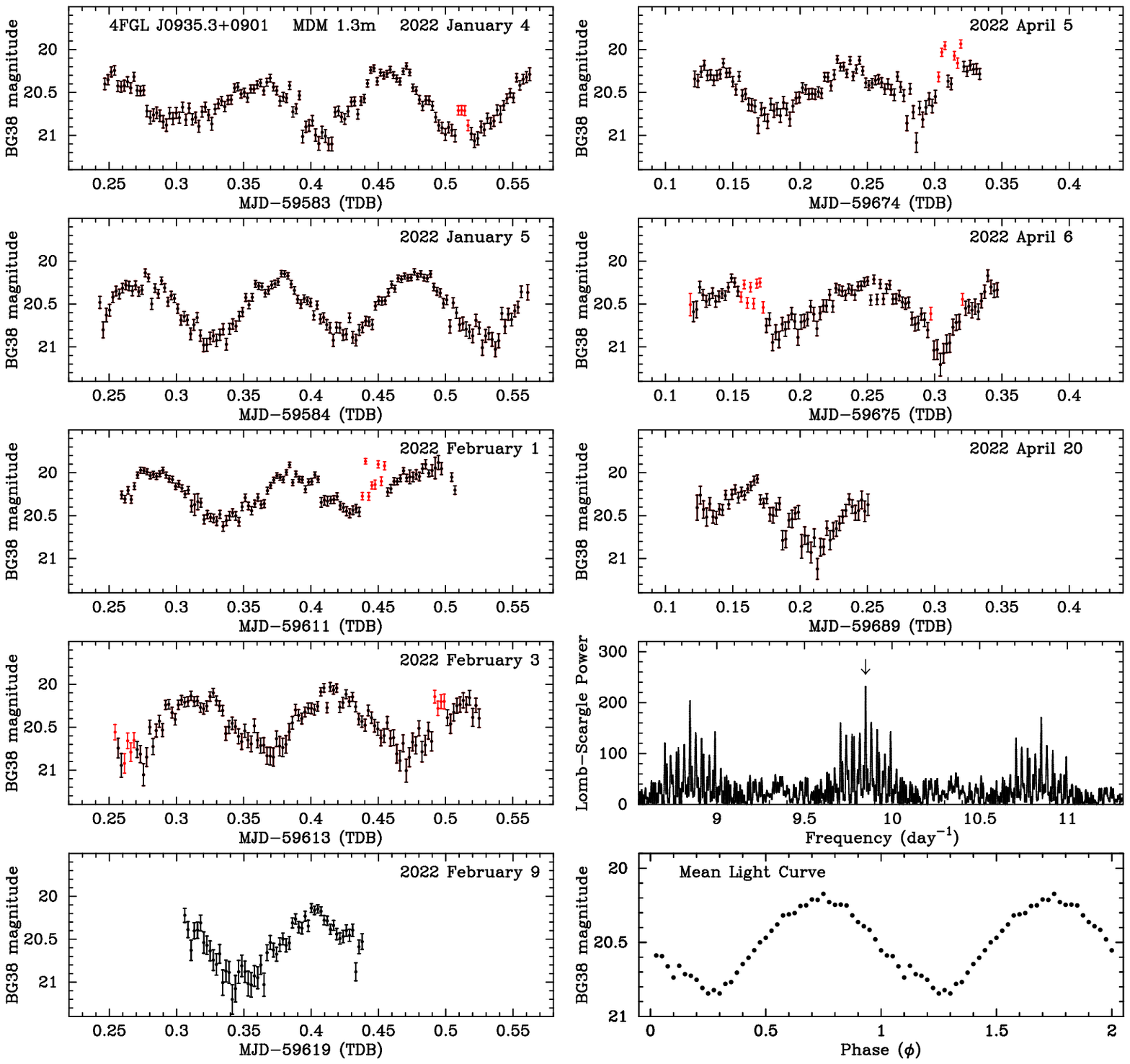}
}
\vspace{-3.1in}
\caption
    {Time-series photometry of \psr\ on the MDM 1.3~m, and a Lomb-Scargle periodogram of the combined data.  The best-fitted period corresponds to the highest peak at 9.85~cycles~day$^{-1}$ (arrow).  Points in red were excluded from the periodogram and the mean folded light curve.}
\label{fig:mdm}
\end{figure*}

A log of the observations is given in Table~\ref{tab:optlog}.  Figure~\ref{fig:mdm} shows the eight nights of photometry, and a Lomb-Scargle periodogram of the combined nights that identifies the period (listed in line 1 of Table~\ref{tab:ephem}).  The time system used is Barycentric Dynamical Time (TDB), as implemented by \citet{eas10}.  Broad modulation of up to 1~mag on the 2.44~hr period is seen, resembling the heating light curves of redback and black widow companions.  Its amplitude is too large to be due to tidal distortion of a Roche-lobe filling companion.  There is also significant night-to-night variation of the amplitude and shape of the light curve.

On several nights, rapid flaring is evident on the time scale of the 200~s exposures.  In order to mitigate random effects, I excluded the more obvious flares, shown as red points in Figure~\ref{fig:mdm}, from the timing analysis.  But this not a perfect filter; it is possible that continuous flaring was responsible for the increased average brightness of some of the observations, e.g., on 2022 February~1.

The mean folded light curve is also shown in Figure~\ref{fig:mdm}.  Phase $\phi=0$ is defined here as the ascending node of the putative pulsar, assuming $\phi=0.25$ (minimum light) is the inferior conjunction of the companion and $\phi=0.75$ (maximum light) is the superior conjunction of the companion.   But the light curve is evidently not precisely symmetric, e.g., the maxima and minima are not exactly 0.5 cycles apart.  So there may be a systematic uncertainty of up to 0.05 cycles (0.005~days) in the correspondence of phase to a kinematic ephemeris.

\begin{deluxetable*}{lccll}
\label{tab:ephem}
\tablecolumns{5} 
\tablewidth{0pt} 
\tablecaption{Photometric Orbital Ephemeris of \psr}
\tablehead{
  \colhead{Source} & \colhead{Method\tablenotemark{a}} & \colhead{Time Span}
  & \colhead{$T_o$\tablenotemark{b}} & \colhead{$P_{\rm orb}$} \\
 & & \colhead{(MJD)} & \colhead{(MJD TDB)} & \colhead{(day)}
}
\startdata
MDM      &  LS    &  59583--59689  &  59584.4031(24) &  0.1015357(37) \\
MDM      &  TOA   &  59583--59675  &  59584.4012(9)  &  0.1015339(22) \\
ZTF      &  LS    &  58217--59550  &  58850.015(5)   &  0.1015333(10) \\
\hline
MDM+ZTF  &  TOA   &  58217--59675  &  59584.4015(7)  &  0.10153276(36) \\
\enddata
\tablenotetext{a}{LS --- Lomb-Scargle periodogram; TOA --- fitting times of arrival of extrema.}
\tablenotetext{b}{Epoch of phase zero, the presumed ascending node of the pulsar.}
\end{deluxetable*}

I also employed a second method to derive an ephemeris: fitting the times of minimum and maximum that could be measured from relatively smooth portions of the light curve.  A total of 16 such times of arrival (TOAs) were fitted to a constant period.  The result is listed in line 2 of Table~\ref{tab:ephem}.  It is consistent with the result of the Lomb-Scargle analysis, but with a smaller uncertainty.  The improvement undoubtedly comes from avoiding flaring by choosing well-behaved portions of the light curve.  The typical residuals of the TOAs from the fit are $\sim200$~s, comparable to the individual exposure times, although a couple of residuals are twice as large as this.  As the residuals are $<0.05$ orbital cycles, it is easy to establish and maintain a phase-connected ephemeris.

\subsection{Zwicky Transient Facility\label{sec:ztf}}

\begin{figure}
  \centerline{
\includegraphics[angle=270.,width=1.0\linewidth]{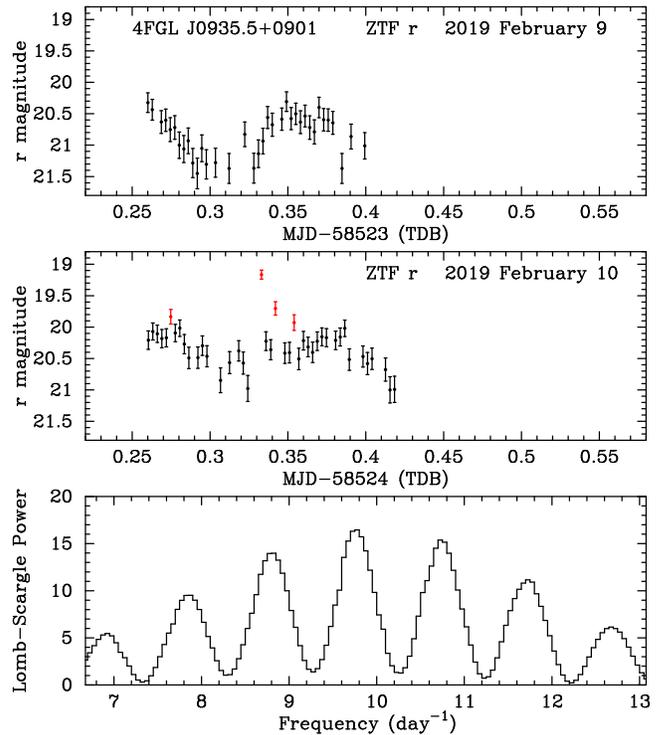}
  }
  \caption
{High-cadence observations by ZTF on two consecutive nights in 2019.  Note the overall change of $\approx0.4$~mag between nights, as well apparent flaring up to $r=19.16$ on the second night (red points).  The bottom panel shows the Lomb-Scargle periodogram of the combined nights' data, excluding points having $r<20.0$.}
\label{fig:ztf}
\end{figure}

The 10th ZTF data release has 195 good exposures of \psr\ in the $r$ band, and 48 good exposures in the $g$ band, over the years 2018--2021.  I start with a pair of high-cadence series in $r$ on two consecutive nights in 2019 February, each covering more than one orbit. Together, these comprise 79 data points.  Shown in Figure~\ref{fig:ztf}, they have the same characteristics as the MDM data, namely, large variation between nights, and fast flares.  After excluding obvious flare points having $r<20.0$, the joint periodogram of the two nights selects the same period as the MDM observations.

\begin{figure}
  \centerline{
\includegraphics[angle=270.,width=1.0\linewidth]{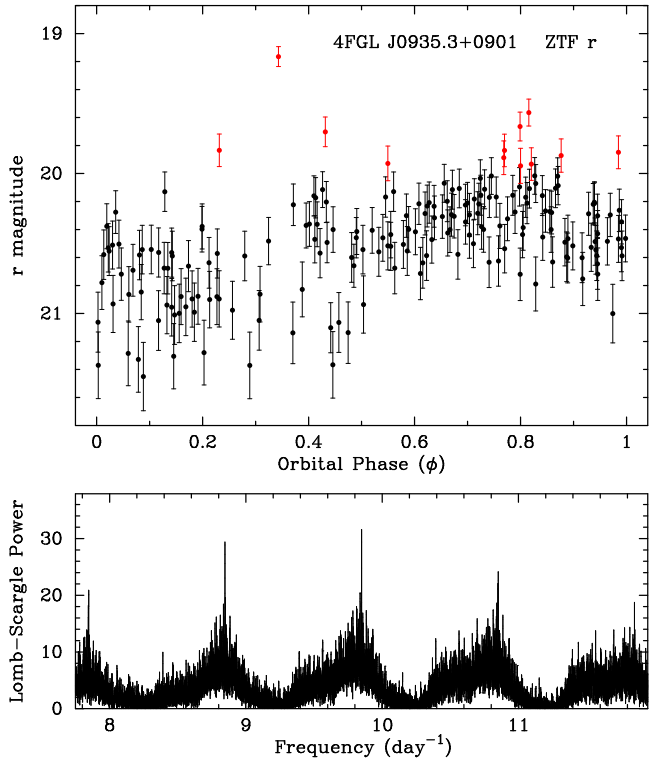}
  }
  \caption
{Bottom panel: a periodogram of the entire ZTF $r$-band data set, but excluding points with $r<20.0$ (red, top), identifies the strongest peak as 9.85~cycles~day$^{-1}$.  The corresponding ephemeris is listed in line 3 of Table~\ref{tab:ephem}.  Top panel: folded light curve with phase defined such that maximum light occurs at $\phi=0.75$, the presumed superior conjunction of the companion star.}
\label{fig:ztf_fold}
\end{figure}

The remaining 60\% of the $r$-band data points spread over $\approx3.6$~yr also contribute to a coherent signal at the same period.  A periodogram and folded light curve of the entire ZTF $r$-band data set is shown in Figure~\ref{fig:ztf_fold}.  It is clear from the fold that \psr\ is reliably detected at maximum light $(\phi=0.75$), while its minimum is close to the detection limit of ZTF.  (The same effect is seen in the $g$ band, albeit with fewer data points.) With flaring being frequent at all orbital phases, exclusion of the brightest flares with $r<20.0$ helps to reduce the noise and identify the true period among the 1 day aliases.  The best-fitted period from this analysis in given in line 3 of Table~\ref{tab:ephem}.  Because of the long span of the data, the ZTF period is twice as precise as the one from the 3.5~month monitoring, albeit denser, at MDM.  However, the ZTF epoch of phase zero $T_0$ is less precise.

\subsection{Joint MDM/ZTF Analysis}

Finally, it is possible to combine the data from both observatories to further refine the ephemeris.  Here, the superiority of TOA fitting for this star recommends itself for the joint analysis.  However, the sparse ZTF data cannot match the precision of the 16 MDM TOAs, created with longer exposures and higher cadence, even for the dense ZTF light curves of 2019 February in Figure~\ref{fig:ztf}.  So I chose instead to create a single TOA from the folded $r$-band ZTF data, representing the midtime around MJD 58,850.  Assigning the time of maximum light to $\phi=0.75$ defines the TOA.

Initial projections of the cycle count using the independent MDM and ZTF derived periods then show that the ZTF TOA falls within 0.1 orbital cycles of the predicted time, supporting the assumption that the cycle count is maintained.  A joint fit of a constant period to the 16+1 TOAs then is used to iteratively adjust the ZTF period and TOA, converging to the final, joint ephemeris in line 4 of Table~\ref{tab:ephem}.

The assumption of a constant period over the 4~yr span of these observations is justified by the level of  precision of the TOAs.  Although orbital periods of black widows and redbacks are known to vary on this timescale, the effect is only detectable through pulsar timing, where it becomes evident as a drift of the time of ascending node by as much as several tens of seconds for redbacks, e.g., \citet{den16}.  Such a small deviation is not within the precision of our TOAs to detect.


\section{Discussion\label{sec:disc}}

\subsection{Redback or Black Widow?}
\label{sec:comp}
In the absence of either pulsar timing, spectroscopy of the companion's photosphere, or colors of the photosphere, only the 2.44~hr orbital period of \psr\ can be used to guess whether the companion has a mass in the range of redbacks or black widows.  Although there is a great deal of overlap, the orbital periods of the 24 redbacks and redback candidates listed in \citet{str19} are all $>3.5$~hr, while 6 of 30 black widows have periods $<2.4$~hr (Figure~6 of \citealt{ray22}).  Therefore, it is a priori more likely that \psr\ is a black widow.

On the other hand, the $\approx1$~mag amplitude of modulation of \psr\ is similar to that of redbacks, and smaller than that of most black widows, which are intrinsically fainter stars, and are difficult to detect on their ''night'' sides.  Minimum light of a black widow companion is usually 2--4 mag fainter than its maximum \citep{dra19}.  Of the nine black widows studied by \citet{dra19}, only one, PSR~J0636+5128, has a light curve with an amplitude as small as 1~mag.  Modeling of the light curve in this case naturally fitted a small inclination angle of $23^{\circ}$.  The same could explain the light curve of \psr.

Independent evidence of the nature of the companion may come from the composition of its photosphere or stellar wind.  The emission lines of the flaring redbacks \msptwo\ (Figure~8 of \citealt{str19}) and \obj\ \citep{hal22} are dominated by hydrogen.  But black widow companions that have mass $\le0.05\,M_{\odot}$ have been stripped to a core where hydrogen will have been depleted, at least to some extent \citep{ben13}.  This is certainly true for the flaring black widow PSR~J1311$-$3430 \citep{rom15}, where hydrogen is absent from both emission lines and photospheric absorption.  However, PSR~J1311$-$3430 is a very small system, with an orbital period of 1.56~hr and a companion mass of only $0.01\,M_{\odot}$.  On the opposite end of the black widow mass range, PSR~J1555$-$2908, with an orbital period of 5.6~hr and a companion mass of $0.06\,M_{\odot}$, has a photospheric spectrum dominated by hydrogen lines \citep{ken22}.
The spectrum of \psr\ contains both H and \ion{He}{1} emission.  While it is not immediately obvious if their strengths indicate unusual abundances, it appears from Figure~5 of \citet{wan20} that the \ion{He}{1} emission lines in \psr\ are stronger relative to hydrogen than they are in the redbacks \msptwo\ and \obj.  Thus, it seems plausible that \psr\ has a hydrogen-deficient atmosphere that would favor a black widow classification

\subsection{Origin of Flare Phenomena}

A minority of redbacks and black widows have optical emission lines, but when detected the lines are highly variable, as is the optical continuum.  The principal examples of this phenomenon are the redbacks \fgl, \msptwo, and \obj\ \citep{hal17b,cho18,str19,mir21,hal22}, and the black widow PSR~J1311$-$3430 \citep{rom15,an17}.  The same objects also have very luminous X-ray flares shown in the above references, as well as lower-level X-ray variability not restricted in orbital phase.  The latter behavior may also describe the \xmm\ observation of \psr\ by \citet{zhe22}.  Although highly variable, \psr\ does not display the discrete moding behavior of tMSPs, which led \citet{zhe22} to conclude that it is in a nonaccreting state.  The optical variability of \psr\ supports this interpretation.

Flares correlated with broad optical emission lines implicate the stellar wind as being involved in both.  In \obj, contemporaneous photometry and spectroscopy show that emission-line strength is correlated with the flaring state \citep{hal22}.  Shocks between the pulsar wind and the stellar wind could accelerate synchrotron-emitting electrons, while the same shocks could ionize and heat the stellar wind to produce the emission lines.  The particle acceleration could be either diffusive or, more likely, powered by magnetic reconnection in the striped pulsar wind that is compressed by the shock \citep{cor22}.  See \citet{aln18} for a review of this subject.

Some spiders have flares and emission lines, while others are steady sources (save for orbital modulation) and have no emission lines.  It is not clear that the difference is simply a function of the spin-down power of the pulsar or the strength of the stellar wind.  Most of the flaring ones (\obj, \fgl, \psr) do not even have pulsations detected, so their spin-down power is unknown.  It may be that a combination of properties of the pulsar and the companion controls whether they will flare.

\subsection{Implications of Radio Nondetection}

\citet{zhe22} conducted a 20 minute pulsar search of \psr\ with FAST at 1.05--1.45~GHz, which did not result in a detection.  The orbital ephemeris of \psr\ can now be used to determine the phase at which the FAST observation was made.  It took place on 2020 August 23 12:35--12:55 UTC, corresponding to MJD 59084.5194--59084.5333 TDB.  Using the MDM+ZTF ephemeris of Table~\ref{tab:ephem}, this corresponds to $0.64\le\phi\le0.78$.  These orbital phases span a superior conjunction of the companion star, the ideal configuration for seeing unabsorbed radio emission from the pulsar.

Similar candidates are often not detected in radio, sometimes due to absorption by the ablated stellar wind at an unfavorable orbital phase.  However, there are a number of new redback candidates, e.g., 4FGL~J2331.1$-$5527 \citep{swi20}, 3FGL~J0940.6$-$7609 \citep{swi21}, and 4FGL J1702.7-5655 \citep{corb22}, that were previously searched a number of times without success \citep{cam15,cam16}. It could be that they are simply beamed away from Earth.   It is also possible that they are sometimes surrounded by a stellar wind that variably wraps around the pulsar even at normally favorable phases \citep{tav93}.  In an extreme case, the mass and energy flux of the evaporated wind may enough to completely enshroud the pulsar and absorb the radio signal at all orbital phases \citep{tav91}.  It would be interesting if such effects are correlated with the strength of emission lines and flaring, properties that could predict transitory breaks in the wind. In addition, if the inclination angle is as small as argued in Section~\ref{sec:comp}, the wind may be less effective as an absorber of the radio.

\section{Conclusions and Suggestions for Further Work}

\psr\ belongs to a small subgroup of spider pulsars, including both redbacks and black widows, that have a variable heating light curve, fast flaring, and optical emission lines.   This cluster of properties can be interpreted as the product of shocks between the relativistic pulsar wind and the stellar wind.  With an orbital period of 2.44~hr and a light curve of modest ($\sim1$~mag) amplitude, it is not immediately obvious whether the companion star falls among the redback or black widow classes, which are defined primarily by mass.  The orbital period favors a black widow, with a small inclination angle to explain the weak orbital modulation.

A future radial velocity study of the companion star, assuming that its photospheric absorption-line spectrum is accessible, would provide a lower limit on the mass of the putative neutron star.   Complementary information could come from multiband light curves that measure the intrinsic temperature of the companion, the external heating from the pulsar, and the inclination angle of the system.  Multiband light curves should ideally be simultaneous to mitigate variability, which is often seen in \psr\ on a timescale of one orbit or less.  Multiband detection of fast optical flares may also reveal their thermal or nonthermal nature.

The phase-connected ephemeris presented here spans 4~yr, providing a precise orbital period and a predicted epoch of ascending node, although the latter parameter is somewhat uncertain because of irregular variability and slight asymmetry of the mean light curve.  This ephemeris will be useful for choosing the orbital phases to observe in any new radio pulsar search.  Higher radio frequencies may be preferred in order to penetrate any absorbing wind.  Searches of archival Fermi $\gamma$-ray data are also possible, and have occasionally discovered pulsations from spiders \citep{ple12,nie20,cla21} with the help of the position and orbital parameters obtained optically.

\begin{acknowledgements}

I thank John Thorstensen for contributing a crucial observation on 2022 February 9, and Slavko Bogdanov for useful discussions.

\end{acknowledgements}

\end{document}